# Electromagnetic Form Factors and the Proton Radius Puzzle: A Critical Review of Extraction Methods via Electron-Proton Scattering

**Gerome A. Paterez[1*], Jade C. Jusoy[1], Edmar Pantohan[1], and Eulogio Auxtero Jr.[1]**
1 Department of Physics, Caraga State University, 8600, Ampayon, Butuan City, Agusan del Norte, Philippines

E-mail: gerome.paterez@carsu.edu.ph

**Abstract**
The study of protons, fundamental constituents of atomic nuclei, is crucial for advancing our understanding of nuclear physics and the fundamental forces of nature. This review paper examines the advancements in the extraction of electromagnetic form factors, which are essential for understanding the internal structure of protons. It highlights the significance of electron-proton scattering experiments, particularly those conducted at facilities like Jefferson Lab (JLab) and Mainz Microtron (MAMI) [2, 3]. The paper discusses various methodologies, including the Rosenbluth separation and polarization transfer techniques, emphasizing their evolution and the challenges faced, such as systematic uncertainties and theoretical ambiguities [23]. The results from the literature indicate that the Rosenbluth separation method yielded a proton radius of approximately 0.8 fm, consistent with earlier measurements but not aligning with more recent polarization transfer results, which suggest a smaller radius [20]. The polarization transfer method has emerged as the most reliable approach, providing more accurate measurements of the ratio of electric to magnetic form factors, particularly in high momentum transfer regions [13, 23]. Additionally, alternative methods like cross-section ratios have been utilized to extend form factor studies, although they still encounter residual uncertainties [3]. The paper concludes that while significant progress has been made in measuring proton form factors, substantial discrepancies remain, particularly in proton radius extraction. Future research is essential to reduce systematic uncertainties and enhance our understanding of the proton's electromagnetic properties, marking a promising direction for advancements in particle physics [2, 23].

## 1. Introduction

In 1917, Ernest Rutherford identified the proton during experiments on alpha particle scattering by thin metal foils. This discovery initiated detailed studies of subatomic particle structure. Protons are now understood to be composite particles composed of quarks and gluons [1]. Quarks are the fundamental constituents of protons and neutrons, while gluons mediate the strong nuclear force that binds quarks together. Quantum Chromodynamics (QCD) provides the theoretical framework to describe these interactions [2]. It is the fundamental theory describing the interactions of quarks and gluons, through the strong nuclear force. It accounts for key phenomena such as confinement, where quarks cannot exist independently, and asymptotic freedom, where quarks interact weakly at high energies, forming the basis for understanding strongly interacting matter.

One of the issues in proton studies is the proton radius puzzle, which arises from discrepancies in radius measurements obtained via different experimental methods. Resolving this issue is critical for advancing the understanding of proton structure and its interactions [2, 3, 4, 5]. Understanding the structure of the proton is fundamental to modern physics as it bridges quantum chromodynamics (QCD) and electroweak interactions, providing critical insights into the behavior of strongly interacting matter.

A key focus in this pursuit is the study of electromagnetic form factors, which encode the spatial and momentum distributions of the proton's charge and magnetization [6, 7]. These form factors are essential for testing theoretical models of nucleon structure and serve as benchmarks for probing deviations from the Standard Model [8, 9].

Electromagnetic form factors are extracted from electron-proton scattering experiments and Hydrogen Spectroscopy, where precision measurements of cross-sections and polarization observables are analyzed [10, 11]. Despite significant progress, extracting these form factors remains challenging due to experimental uncertainties, model-dependent assumptions, and radiative corrections [12]. These difficulties intensify when estimating the proton charge radius, linked to the slope of the electric form factor at zero momentum transfer. The electromagnetic form factor technique efficiently addresses the proton radius puzzle using precise electron-proton scattering, directly probing the proton's charge distribution with minimal systematic uncertainties. Unlike muonic hydrogen spectroscopy, it avoids dependence on complex atomic energy level calculations. High-precision setups, like PRad, further reduce extrapolation errors [13, 14]. The proton radius puzzle emerged in 2010 by observing discrepancies between the proton radius obtained from muonic hydrogen spectroscopy and electron-proton scattering experiments [15, 16]. This inconsistency has spurred debates, questioning the universality of electromagnetic interactions and the reliability of extraction techniques [17, 18]. Resolving this puzzle is critical for refining our understanding of proton properties and ensuring consistency across physics domains.

This review examines methods for extracting electromagnetic form factors and the proton radius through electron-proton scattering, analyze underlying assumptions, methodologies, and sources of systematic uncertainty influencing these measurements. Additionally, this paper explores whether current discrepancies arise from methodological biases or suggest new physics.

Electron-proton scattering experiments, including Rosenbluth separation and polarization transfer techniques, have been the primary methods for extracting these quantities [11, 14]. Studies conducted at facilities such as Jefferson Lab (JLab) and Mainz Microtron (MAMI) have provided extensive data sets for analyzing nucleon structure. The determination of form factors faces significant challenges due to experimental uncertainties, model-dependent corrections, and theoretical ambiguities. Inconsistencies between results from different techniques, such as discrepancies in the ratio of electric to magnetic form factors, highlight these issues [15, 16]. The incorporation of two-photon exchange effects has improved theoretical models, but substantial uncertainties remain in addressing these corrections [12].

The proton radius puzzle, arising from a discrepancy between radii determined via muonic hydrogen spectroscopy and electron-proton scattering, remains unresolved [13, 14]. This issue has led to debates on the reliability of extraction methods and the potential need for new physics. Ongoing efforts to refine experimental precision and improve theoretical frameworks are crucial for addressing these inconsistencies [15, 16].

Two primary processes can occur in electron-proton scattering: elastic scattering and inelastic scattering. At low energies, the dominant process is elastic scattering, where the proton remains intact. Elastic scattering is described by the coherent interaction of a virtual photon with the proton as a whole. It thus provides a probe of the global properties of the proton, such as its charge radius. At high energies, the dominant process is deep inelastic scattering, where the proton breaks up. Here, the underlying process is the elastic scattering of the electron from one of the quarks within the proton. Consequently, deep inelastic scattering provides a probe of the momentum distribution of the quarks [17, 8].

Given that the electron is one of the elementary particles, it is a significantly light particle that can penetrate the nucleus during the scattering process. However, during the scattering, the electron does not directly undergo a smooth scattering to the nucleus because the nuclear magnetic field couples it since the proton has a non-zero spin [19, 20]. When conducting the electron-proton scattering, it is significant to determine the cross-section.

$$dcs_{Mott} = \left(\frac{d\sigma}{d\Omega}\right)_{Mott} = \left(\frac{(Z_e Z_p \alpha)^2}{4k sin^4(\theta/2)}\right)\left(\frac{E'}{E}\right)(1 - v^2 \sin^2(\theta/2)) \quad (1)$$

Equation (1) describes the interaction of the electron and proton during the scattering, where the cross-section is denoted by $dcs_{Mott}$ for Mott scattering, $Z_e$ and $Z_p$ are the charges of electron and proton respectively, $\theta$ is the scattering angle, $\alpha$ is the fine structure constant that is approximately has a value of 1/137, $k$ is the incident electron's momentum, $E'$ and $E$ are the final and initial energy of the electron respectively. It is also stated from the equation the ratio of the final and the initial energy of the electron is called the relativistic recoil factor, see Equation (2).

$$\frac{E'}{E} = \frac{1}{1+\left(\frac{2E}{M_p}\right)\sin^2(\theta/2)} \quad (2)$$

The difference between the initial and final energy can be expressed by Equation (3)

$$E - E' = -\frac{q^2}{2M_p} = \frac{Q^2}{2M_p} \quad (3)$$

where $M_p$ is the mass of the proton. For the electron to probe the proton, the electron must acquire enough energy that is the difference between the initial and the final energy, where Equation (4) is the momentum transfer.

$$Q = -q \quad (4)$$

Momentum transfer, $Q$, is a measure of how much momentum is exchanged between the electron and the proton during a scattering event, with larger $Q$allowing the electron to "zoom in" and see finer details of the proton's structure. The four-momentum transfer, $q$, is the difference between the electron's initial and final momenta, $q = k - k'$, and its squared value is often negative in the math framework used (Minkowski space). To make $Q$ positive and easier to interpret as a measure of spatial resolution, it is defined as $Q = -q$. This convention is widely used in experiments to study the proton's size and internal charge distribution [14, 16, 18].

Elastic electron-proton scattering is a fundamental process studied in particle and hadron physics, particularly in understanding the structure of protons and other hadrons. In this process, an electron is scattered off a proton, and both particles emerge from the interaction intact. "Elastic" refers to the conservation of kinetic energy and momentum in the interaction, meaning that no energy is produced or destroyed during the process, and the scattering occurs without any internal changes to the particles involved [23, 24, 25]. During this process, the virtual photon as a force carrier will be used to probe the proton's structure; this type of scattering does not destroy the proton [26]. The proton is not a point-like particle based on recent experiments, therefore it is necessary to take into account that it is built out of constituents and that it has an extension or is not a point anymore. One way to prove this is to get the form factors by analyzing the Fourier Transform of the charge density functions; see Equation (5).

$$F(\vec{q}) = \int d^3\vec{r}\, e^{i\vec{q}\cdot\vec{r}} \rho(\vec{r}) \quad (5)$$

$$F(0) = 1 \quad (6)$$

$$dcs_{ff} \approx dcs_{Mott} \cdot |F(\vec{q})|^2 \quad (7)$$

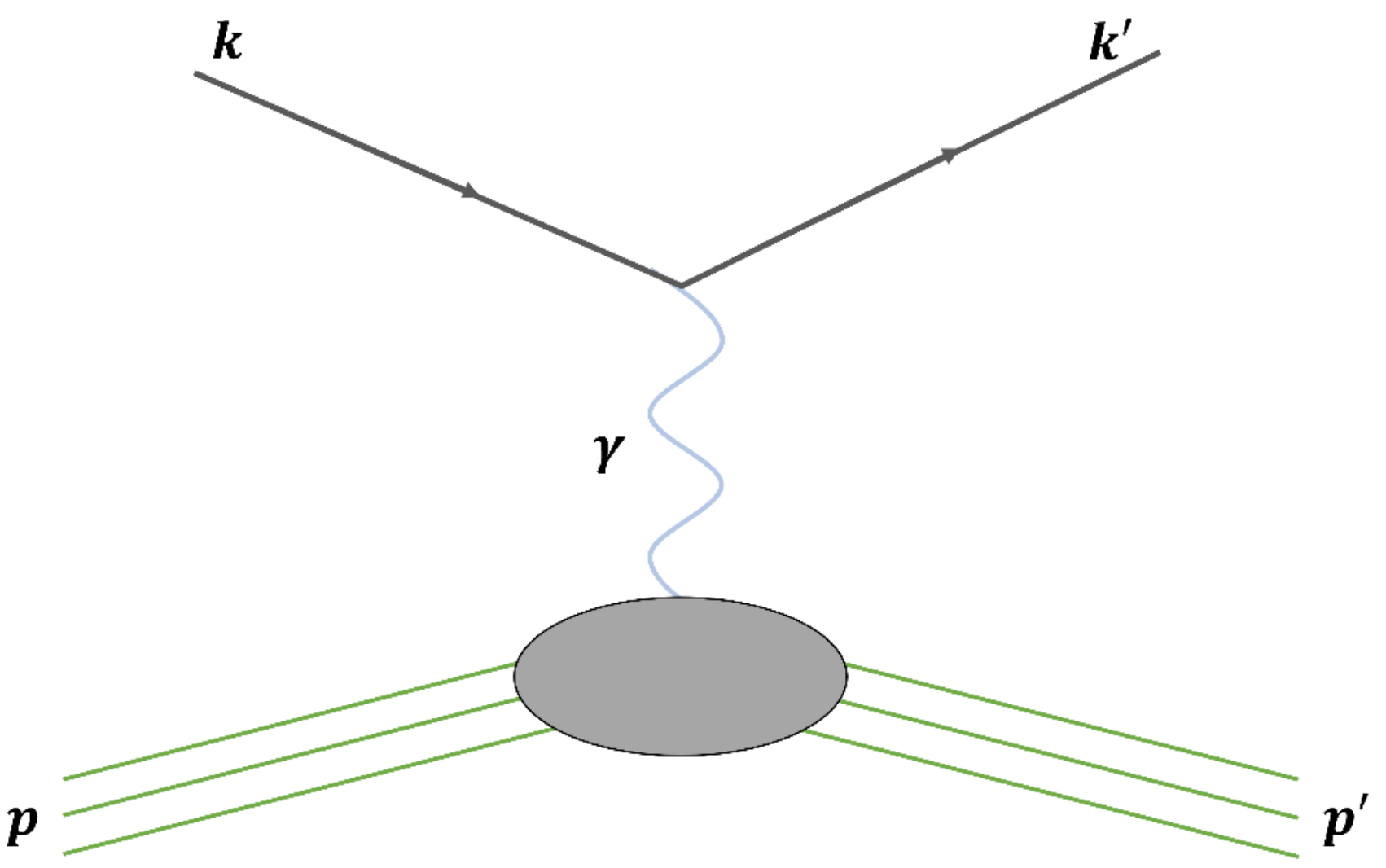

**Figure 1**. Illustrates a schematic diagram of electron-proton scattering. In the diagram, $k$ represents the initial momentum of the incoming electron, $k'$ represents the final momentum of the scattered electron, $p$ represents the initial momentum of the initial proton, and $p'$ represents the final momentum of the scattered proton. The scattering occurs via the exchange of a virtual photon $\gamma$ [25].

Where $\rho(\vec{r})$ is the charge density in real space (typically a function of position $\vec{r}$), $\vec{q}$ is the momentum transfer vector, and the integral is performed over all space, integrating the charge density weighted by a phase factor $e^{i\vec{q}\cdot\vec{r}}$. Equation (7) stated the extension of the cross-section from a point-like cross section via Fourier Transform of the charge distribution. The extended differential cross-section reveals how the proton's internal charge distribution affects its interaction with an electron. This helps to understand how the distribution of the proton's charge affects its interaction with an electron and/or revealing details about how the charge is spread throughout the proton. The Feynman diagram in Figure (1) can be described in mathematical modelling. Starting from the scattering amplitude:

$$\mathcal{M} = \frac{4\pi\alpha}{q^2} J_\mu^{electron}(q) J^{\mu,proton}(q) \qquad (8)$$

The scattering amplitude Equation (8) is a quantity that describes the probability amplitude for a particular scattering process between elementary particles. Its function is to determine the likelihood of specific interactions occurring, such as particle collisions, decays, or exchanges, and it is directly related to observable quantities like cross-sections and decay rates. From the scattering amplitude, three factors need to be considered; two are functions of the energy transfer. First, the leptonic current or the current due to the beam of the electron which is given by:

$$J_\mu^{electron}(q) = -e\bar{u}_e \gamma_\mu u_e(k) \qquad (9)$$

where $k$ and $k'$ represent the initial and final momentum of the electron, respectively. The $\bar{u}_e$ and $u_e$ denote the corresponding Dirac spinors, while $\gamma_\mu$ represents the basic vertex factor for the electron. Meanwhile, the hadronic current for a spin-$\frac{1}{2}$ particle is described by Foldy [24]:

$$J^{\mu proton}(q) = e\bar{u}_N(p_2)\left[F_1(Q^2)\gamma^\mu + F_2(Q^2)\frac{i\sigma^{uv}}{2M_p}q_v\right]u_N(p_1) \qquad (10)$$

The Pauli form factor $F_2$ and the Dirac form factor $F_1$, in the context of one-photon exchange, are real functions that depend exclusively on the squared energy transfer $Q^2$. Importantly, they preserve relativistic invariance. Substituting Equation (9) and Equation (10) to Equation (8) and simplifying, the equation for the scattering amplitude becomes,

$$\mathcal{M} = \frac{e^2}{Q^2}\, e\,\bar{u}_e(k')\gamma_\mu u_e(k)\, u_N(p_2)\left[F_1(Q^2)\gamma^\mu + F_2(Q^2)\frac{i\sigma^{uv}}{2M_p}q_v\right]u_N(p_1) \qquad (11)$$

the cross-section formula for electron-proton (ep) scattering in the laboratory frame is expressed as,

$$dcs = \frac{|\mathcal{M}|^2}{4\left((k\cdot p)^2 - m_e^2 M_p^2\right)} \frac{(2\pi)^4\delta^4(k'-k-p')}{\left((2\pi)^3 2E'\right)\left((2\pi^3)2(M_p+\omega)\right)} \qquad (12)$$

using the scattering amplitude to study proton and electron interactions, the momentum conservation requirement, enforced by a delta function δ, is integrated out. Consequently, the proton current can be represented using the Dirac and Pauli form factors. The differential cross-section can be expressed as:

$$dcs = \frac{\alpha^2}{4E^2\sin^4\left(\frac{\theta_e}{2}\right)}\left(\frac{E'}{E}\right)\left[\left(F_1^2 + \frac{\mu_p^2 Q^2}{4M_p^2}F_2^2\right)\cos^2\left(\frac{\theta}{2}\right) + \frac{Q^2}{4M_p^2}\left(F_1 + \mu_p F_2\right)^2\sin^2\left(\frac{\theta}{2}\right)\right]\delta\left(E - E' - \frac{Q^2}{2M_p}\right) \qquad (13)$$

The integration involves averaging over initial spins and summing over final spins. The delta function ensures elastic scattering in this scenario. In Equation (13), integration is performed, and then both the numerator and denominator are divided by $\cos^2(\theta/2)$.

$$dcs_{ff} = \frac{(Z_e Z_p \alpha)^2}{4k^2\sin^4\left(\frac{\theta}{2}\right)}\left(\frac{E'}{E}\right)\left[1 - v^2\sin^2\left(\frac{\theta}{2}\right)\right] \times \left[\left(F_1^2 +\right.\right.$$

$$\frac{\mu_p^2 Q^2}{4M_p^2}F_2^2\Big) + \frac{Q^2}{4M_p^2}\left(F_1 + \mu_p F_2\right)^2 \tan^2\left(\frac{\theta}{2}\right)\Big] \quad (14)$$

$$dcs_{ff} = dcs_{Mott}\left[\left(F_1^2 + \frac{\mu_p^2 Q^2}{4M_p^2}F_2^2\right) + \frac{Q^2}{4M_p^2}\left(F_1 + \mu_p F_2\right)^2 \tan^2\left(\frac{\theta}{2}\right)\right] \quad (15)$$

The convention is chosen to highlight the fundamental nature of space-time, where distance and time are treated on an equal footing (in units where c=1=h, they have the same units). Therefore, quantities like momentum and energy can be directly related without separate considerations for the speed of light, making equations more compact and intuitive in the context of high-energy particle interactions and relativity. Therefore, by this condition, the Equation (14) reduces to:

$$dcs = \frac{(Z_e Z_p \alpha)^2}{4k^2 \sin^4\left(\frac{\theta}{2}\right)}\left(\frac{E'}{E}\right)\cos^2\left(\frac{\theta}{2}\right)\left[\left(F_1^2 + \frac{\mu_p^2 Q^2}{4M_p^2}F_2^2\right) + \frac{Q^2}{4M_p^2}\left(F_1 + \mu_p F_2\right)^2 \tan^2\left(\frac{\theta}{2}\right)\right] \quad (16)$$

the form factor terms in Equation (16) can be written in terms of Sachs electric and magnetic form factors ($G_E = (Q^2), G_M = (Q^2)$):

$$G_E(Q^2) = F_1(Q^2) - \mu_p \tau F_2(Q^2) \quad (17)$$

$$G_M(Q^2) = F_1 + \mu_p F_2 \quad (18)$$

Where $\tau = \frac{Q^2}{4M_p^2}$. It follows that:

$$dcs_{ff} = \frac{(Z_e Z_p \alpha)^2}{4k^2 \sin^4\left(\frac{\theta}{2}\right)}\left(\frac{E'}{E}\right)\cos^2\left(\frac{\theta}{2}\right)\left[\frac{G_E^2 + G_M^2}{1+\tau} + 2\tau G_M^2 \tan^2\left(\frac{\theta}{2}\right)\right] \quad (19)$$

Hence, this can be simplified to:

$$dcs_{ff} = \sigma_{ns}\frac{1}{1+\tau}\left[G_E^2 + \frac{\tau}{\epsilon}G_M^2\right] \quad (20)$$

Where,

$$\frac{1}{\epsilon} = \left[1 + 2\left(1 + \frac{Q^2}{4M_p^2}\right)\tan^2\left(\frac{\theta}{2}\right)\right] \quad (21)$$

$$\frac{1}{\epsilon} = \left[1 + 2(1+\tau)\tan^2\left(\frac{\theta}{2}\right)\right] \quad (22)$$

Electromagnetic form factors $G_E(Q^2)$ and $G_M(Q^2)$ of extended quantum systems are typically viewed as a method for obtaining insights into the distribution of charge, magnetism, and weak charge by analyzing their Fourier transforms, valid only for a light system such as the proton [26]. Electric form factors are functions that describe the spatial charge distribution of the particle. They provide essential information about the particle, such as the internal structure, that might lead to the next level of understanding of the particles. Magnetic form factors are physical and mathematical functions that describe the magnetic moment distribution. They are almost similar to the electric form factors; they provide new insights into the nearly exact shape deviating from the classical dipole form. The electromagnetic form factors of the proton can be extracted through data analysis in an experimental setup. The mechanism involves several steps such as Peak Region Analysis (defined as the region of the spectrum of the scattered electrons where the cross section is expected to peak), Radiative Tail Analysis (the tail of the spectrum of the scattered electrons due to the emission of photons), Cross Section Calculation (determined by correcting the number of events in the peak region for the radiative tail, the energy loss of the electrons in the target material).

Inside the nucleus is a substructure characterized by distributions of electric charge and magnetic moment. The cross-section of a nucleus is formulated as [27],

$$dcs_{ff} = dcs_{Mott}|F(Q^2)|^2 \quad (23)$$

In the limit where relativistic effects are negligible,

$$F(Q^2) = \int \rho(\vec{r})e^{i\vec{Q}.\vec{r}}d^3\vec{r} \quad (24)$$

The charge distribution of the nucleus is represented by $\rho(\vec{r})$. Transforming this distribution using the Fourier transform produces the electric form factor, $G_E(Q^2)$. Similarly, the Fourier transform of the proton's magnetic moment distribution results in the magnetic form factor, $G_M(Q^2)$. Therefore, obtaining $G_E(Q^2)$ and $G_M(Q^2)$ involves analyzing these factors:

$$G_E(Q^2) = \int \rho(\vec{r})e^{i\vec{Q}.\vec{r}}d^3\vec{r} \quad (25)$$

$$G_M(Q^2) = \int \mu(\vec{r})e^{i\vec{Q}.\vec{r}}d^3\vec{r} \quad (26)$$

Calculating the Fourier transforms of both form factors when $Q^2 = 0$,

$$G_E(0) = \int \rho(\vec{r})d^3\vec{r} = 1 \quad (27)$$

$$G_M(0) = \int \mu(\vec{r})d^3\vec{r} = +2.79 = \mu_p \quad (28)$$

In non-relativistic scenarios, the electric and magnetic form factors are connected in the following way,

$$G_E(Q^2) = \frac{G_M(Q^2)}{\mu_p} \quad (29)$$

## 2.1. Extraction Methods for Electromagnetic Form Factors

### 2.1.1. Rosenbuth Separation

The Rosenbluth separation method relies on measuring the differential cross-section of unpolarized electron-proton scattering at various angles and momentum transfer. By analyzing the dependence of the cross-section on the scattering angle, the electric $G_E(Q^2)$ and magnetic $G_M(Q^2)$ form factors can be separated. However, this technique is sensitive to systematic uncertainties, such as those arising from radiative corrections and two-photon exchange contributions [14].

The theoretical foundation of the Rosenbluth separation method stems from quantum electrodynamics (QED) and the hadronic structure of the proton. The scattering process is modelled by the exchange of virtual photons between the electron and the proton. The form factors$G_E(Q^2)$ and \( $G_{M_p}$ encapsulate the spatial distributions of the proton's charge and magnetization, respectively. The general formalism for unpolarized electron-proton scattering in the lowest order of QED gives rise to the following expression for the differential cross-section:

$$dcs = \frac{\alpha^2}{4E^2 \sin^4(\theta/2)} [G_E(Q^2) + \tau G_{M_p}(Q^2)] \qquad (30)$$

where $\alpha$ is the fine structure constant, $E$is the energy of the incoming electron, $\theta$ is the scattering angle, and $\tau = Q^2/4M_p^2$ is the kinematic variable, with $Q^2$ being the squared four-momentum transfer and $M_p$ the mass of the proton. This expression assumes that the electric and magnetic form factors are independent and are functions only of the momentum transfer $Q^2$, simplifying the extraction of individual form factors by measuring the cross-section at various values of $Q^2$.

Despite its success, the Rosenbluth separation method is sensitive to several sources of systematic uncertainties. One significant challenge arises from radiative corrections, which are higher-order QED effects that modify the cross-section. These corrections must be carefully accounted for to prevent significant discrepancies, especially at higher momentum transfers [14].

Additionally, two-photon exchange contributions, which are not included in the basic Rosenbluth formalism, introduce further uncertainty. These contributions arise from processes in which the electron and proton interact via the exchange of two virtual photons, distorting the measured cross-sections and leading to errors in the extraction of the form factors [28, 29]. These effects are most pronounced at lower values of $Q^2$, necessitating precise modeling and quantification to ensure accurate form factor extraction. Although improvements in the modeling of these corrections have been made, they continue to represent a significant challenge in achieving precise measurements [14].

Thus, while the Rosenbluth separation method remains a fundamental technique for extracting proton form factors, it must be applied with caution. Modern experiments often incorporate additional techniques, such as polarization transfer and higher-order QED corrections, to enhance accuracy and mitigate systematic errors.

### 2.1.2. Polarization Transfer

Polarization transfer experiments involve measuring the polarization of the recoil proton when a polarized electron beam is scattered off a proton target. This method directly provides the ratio $G_E/G_M$ with reduced sensitivity to radiative corrections compared to Rosenbluth separation. Significant advancements in this approach have been made at facilities like Jefferson Lab, which achieved high precision in form factor measurements [11].

Polarization transfer experiments offer an alternative to the Rosenbluth separation method by directly measuring the polarization of the recoil proton in electron-proton scattering, when a polarized electron beam interacts with a polarized proton target. These experiments provide a direct measurement of the ratio

$$\frac{G_E(Q^2)}{G_M(Q^2)} \qquad (31)$$

The ratio of the electric form factor $G_E(Q^2)$ to the magnetic form factor $G_M(Q^2)$. This method reduces sensitivity to radiative corrections, which are a major source of systematic uncertainty in the Rosenbluth separation technique.

The key advantage of polarization transfer is that it involves a simpler theoretical framework for extracting the form factors, particularly in the limit of high momentum transfer. In this approach, the polarization of the recoil proton can be expressed in terms of the polarization transfer coefficients;

$$P^* = \frac{1}{\sqrt{1+\epsilon}} \cdot \frac{G_E(Q^2)}{G_M(Q^2)} \cdot P_e \qquad (32)$$

Where $P^*$ is the polarization of the recoiling proton after the scattering event, $P_e$ is the initial polarization of the incident electrons, and $\epsilon$ is the virtuality or polarization parameter related to the kinematics of the scattering, defined as:

$$\epsilon = \frac{1}{1+2(1+\tau)\tan^2(\theta/2)} \qquad (33)$$

This method is generally performed at facilities with high-energy electron beams, such as Jefferson Lab, which has played a leading role in the development and implementation of this method. The precision measurements conducted at such facilities have led to more accurate determinations of the proton form factors, with reduced uncertainties compared to traditional methods. By exploiting the sensitivity of polarization observables to the form factor ratio, this technique offers a more direct and less model-dependent way of measuring the electric and magnetic form factors.

Despite its advantages, polarization transfer is not without challenges. One key issue is the need for a highly polarized proton target and an accurate understanding of the reaction dynamics in order to interpret the polarization observables correctly. Additionally, the extraction of individual form factors from the polarization transfer ratio requires careful analysis of the experimental setup and model-dependent assumptions.

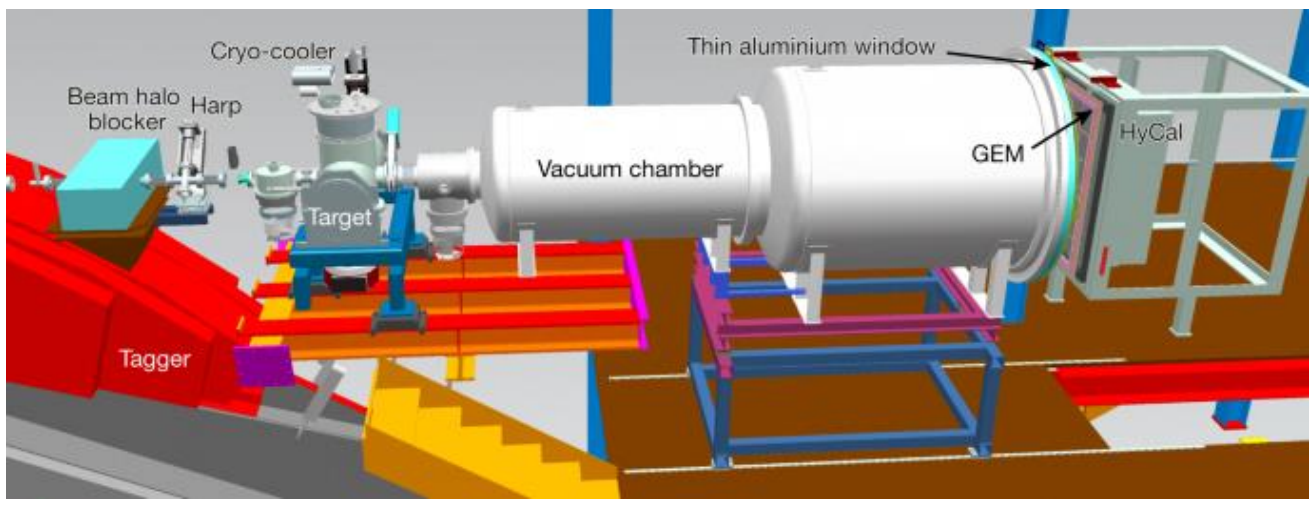

**Figure 2**. Shows the experimental set-up for electron scattering at Jefferson Laboratory [30].

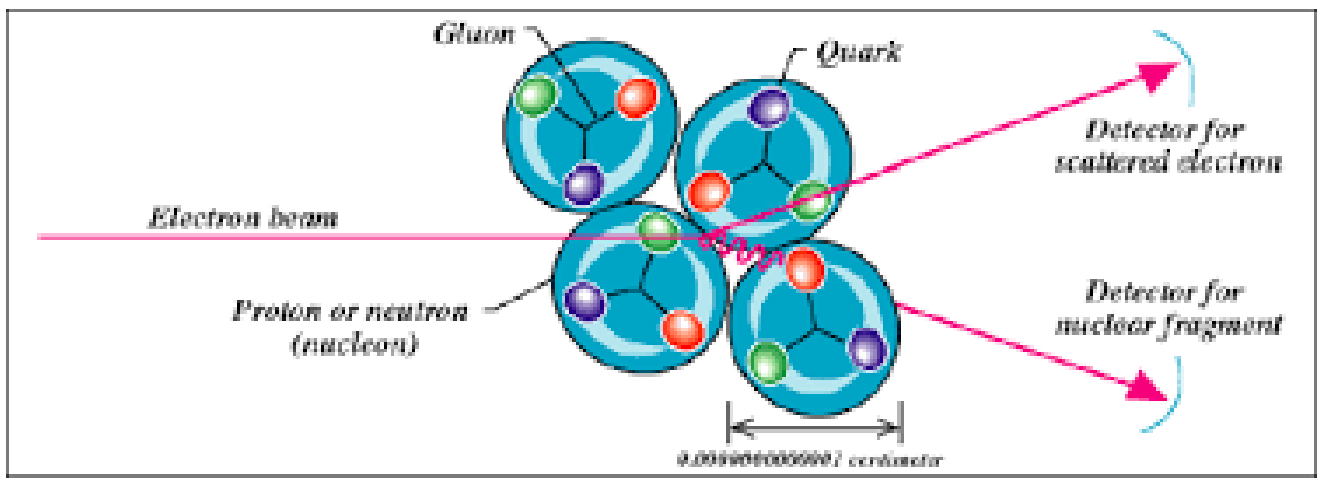

**Figure 3**. Illustrates a schematic diagram of electron scattering interaction experiments by the Jefferson Laboratory [30].

However, with advancements in experimental techniques and an improved understanding of the reaction mechanisms,

polarization transfer experiments have become a powerful tool for form factor extraction, particularly in the high $Q^2$ regime, where Rosenbluth separation is less reliable. Recent work at Jefferson Lab and other facilities has further refined this method, providing more precise data that helps resolve the discrepancies observed in the proton form factor measurements, especially the proton radius puzzle.

### 2.1.3. Cross-Section Ratios

An alternative method involves analyzing the ratio of cross-sections for elastic scattering and other reference processes. This approach minimizes the reliance on absolute cross-section measurements, thereby reducing some systematic uncertainties. It has been particularly useful for extending form factor studies to higher momentum transfer regions [31].

Cross-section ratios allow for the extraction of proton form factors by comparing the elastic scattering cross-section to that of a reference process, such as inelastic scattering or Compton scattering. The ratio is expressed as:

$$\frac{\sigma_{el}(Q^2)}{\sigma_{ref}(Q^2)} = \left(\frac{G_E(Q^2)}{G_M(Q^2)}\right)^2 \cdot F(Q^2) \quad (34)$$

Here, $\sigma_{el}(Q^2)$ and $\sigma_{ref}(Q^2)$ represent the elastic and reference process cross-sections, respectively, while $G_E(Q^2)$ and $G_M(Q^2)$ are the electric and magnetic form factors, and $F(Q^2)$ is a kinematic factor accounting for the differences in dynamics between the elastic and reference processes. This method reduces dependence on absolute cross-section measurements, minimizing systematic uncertainties, and is particularly effective at higher momentum transfers, where other methods are more prone to radiative corrections and two-photon exchange effects [31, 32].

However, challenges persist in accurately modeling the reference process's cross-section. Inaccuracies in this model can introduce substantial uncertainties in the extracted form factors, particularly at high $Q^2$ values where the reference process might be less understood or have complex contributions. Therefore, while the cross-section ratio method offers advantages, careful selection and precise modeling of the reference process are essential to minimize errors.

### 2.1.4. Dispersion Relation

Dispersion relation analyses use the analytic properties of form factors to connect experimental data with theoretical predictions. By incorporating constraints from QCD and low-energy effective theories, this method provides a model-independent framework for interpreting data [33].

This method provides a powerful tool for linking experimental measurements of proton form factors with theoretical models, based on the analytic properties of these form factors. The general idea is that the form factors, such as $G_E(Q^2)$ and $G_M(Q^2)$, can be expressed as functions that satisfy certain analytic constraints. These constraints arise from fundamental principles such as causality and unitarity, and they allow the form factors to be connected to theoretical predictions even at high momentum transfers.

In the context of QCD, dispersion relations incorporate the principles of analyticity, which state that form factors must have a well-defined analytic structure in the complex momentum transfer plane. By using the known behavior of the form factors at large and small momentum transfers, along with low-energy effective field theories like Chiral Perturbation Theory (ChPT), dispersion relations provide a model-independent way to interpolate between experimental data points and make predictions in regions where direct measurements are unavailable.

The theoretical foundation of dispersion relations is rooted in the integral representation of form factors, which relate them to physical data through a Cauchy integral along the contour of the complex plane. These relations take into account both the known asymptotic behavior of the form factors (informed by QCD) and the low-energy expansions provided by effective field theories. By including constraints from both QCD and low-energy effective theories, the method avoids the need for model-dependent assumptions, making it a particularly reliable approach in extracting form factors from experimental data.

The most commonly used dispersion relation for form factors is the Omnès representation, which expresses the form factor as an integral over a known dispersive kernel. This approach is particularly useful for addressing challenges in regions of the momentum transfer where experimental data are sparse, while still ensuring consistency with theoretical models and physical principles [7, 33].

Despite its advantages, the dispersion relation method is not without challenges. One major difficulty is the need for accurate experimental data over a wide range of momentum transfers to ensure that the dispersion integrals converge properly. Additionally, while dispersion relations provide a model-independent framework, they still require careful treatment of the low-energy behavior of form factors, which is influenced by complicated non-perturbative QCD effects. Nonetheless, this method remains a cornerstone in the analysis of form factors, particularly when dealing with high-precision measurements where other methods may suffer from significant systematic uncertainties.

## 2.2. Proton Radius Puzzle

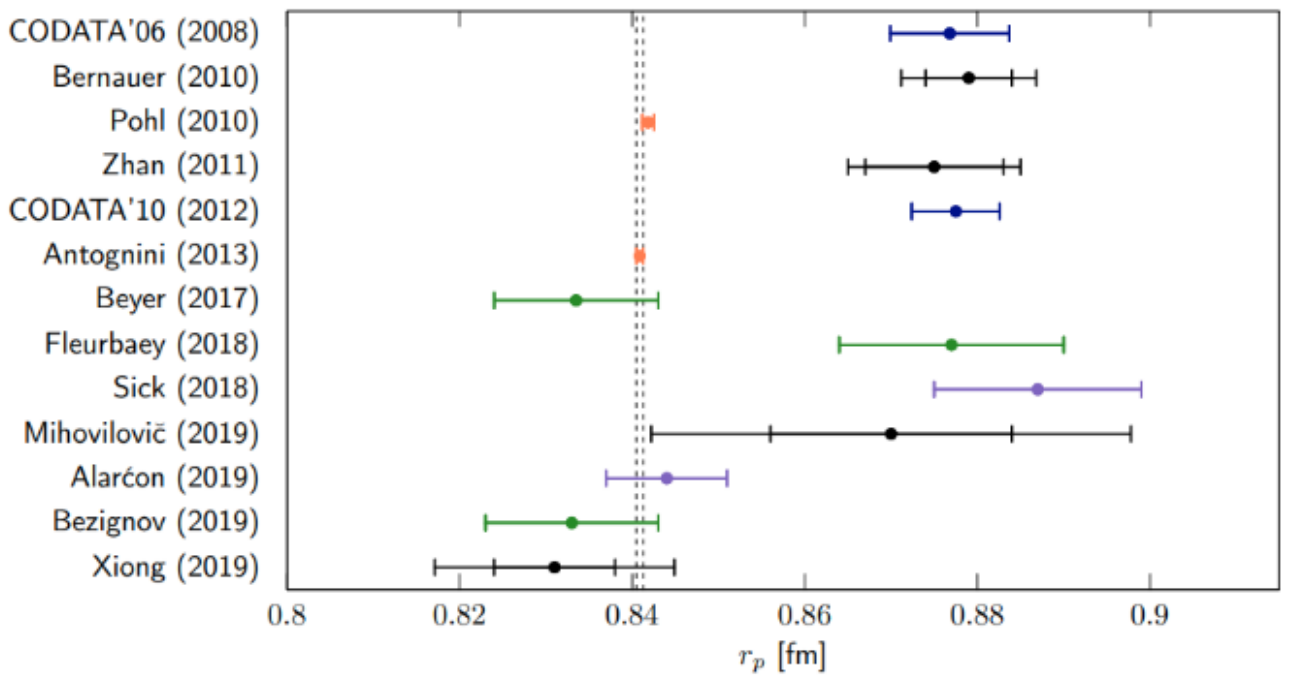

**Figure 4**. The recent report on proton radius with discrepancy indication from different experiments and calculations from 2008-2019. Dark blue points indicate global fits using electron spectroscopy and scattering data. Black points are results from scattering experiments. Purple points are refits of existing data. Green points are from electron spectroscopy results. Orange points are results from muonic spectroscopy experiments [34].

The proton radius puzzle refers to the ongoing discrepancy in the measurement of the proton's charge radius see Figure (4), which has led to significant debate and research within the physics community. The latest findings indicate a trend towards a smaller proton radius, with data suggesting increasingly consistent values across different measurement methods. Notably, integrating muonic hydrogen measurements into the Committee on Data for Science and Technology (CODATA) recommended values marks a pivotal moment in addressing this puzzle. Historically, the proton charge radius was measured by two independent methods, which converged to a value of about 0.877 femtometers [34].

That value was challenged by a 2010 experiment, which produced a radius about 4% smaller at 0.842 femtometers. New experimental results reported in the autumn of 2019 agree with the smaller measurement, as does a re-analysis of older data published in 2022. While some believe this difference has been resolved, this opinion is not yet universally held [4, 33, 36, 37].

The Mainz Microtron (2010) and PRad (2019) experiments demonstrate methodological differences in proton radius measurements. Mainz employed three spectrometers to detect scattered electrons using electron beam energies of 180-855 MeV and covered a $Q^2$ range from 0.004 to 1 $\mathrm{GeV}^2$, resulting in a proton radius measurement of $r_p = 0.879 \pm 0.008$ fm [31]. In contrast, PRad innovated with a magnetic-field-free setup using a windowless hydrogen gas target and novel calorimeter detection system, operating at higher beam energies of 1.1 and 2.2 GeV, and most notably achieving an unprecedented low $Q^2$ range down to $2.1 \times 10^{-4}\ \mathrm{GeV}^2$, which led to a smaller proton radius measurement of $r_p = 0.831 \pm 0.007_{stat} \pm 0.012_{sys}$ fm [38]. This significant difference in experimental setup and momentum transfer range accessibility between the two experiments contributed to their differing results, highlighting the importance of low $Q^2$ measurements in proton radius determinations.

## 3. Results and Discussion

### 3.1. Rosenbluth Separation Method

Walker et al. (1994) applied this method and found that the extracted proton radius was approximately 0.8 fm, which was consistent with earlier measurements from the Rosenbluth separation technique. However, this value did not fully align with the more recent, more precise measurements obtained through polarization transfer experiments [11]. Despite its utility, the Rosenbluth separation method faced significant challenges, primarily from radiative corrections and two-photon exchange contributions. These systematic uncertainties become particularly pronounced at higher momentum transfers, where the method's accuracy is compromised, leading to discrepancies in the extracted form factors [31]. As a result, the method was gradually superseded by the polarization transfer method, which provides more accurate and reliable measurements of the proton form factors [39].

While the Rosenbluth separation method was instrumental in the early exploration of proton form factors, its limitations have made it less preferable in contemporary research. The increasing complexity of corrections required, especially at larger values of $Q^2$, has prompted the use of polarization transfer experiments, which are less sensitive to these uncertainties and offer higher precision, particularly in determining the proton radius.

### 3.2. Polarization Transfer Method

Polarization transfer experiments, such as those by Jones et al. (2000), have become the preferred method for measuring the proton radius due to their precision and reduced sensitivity to systematic uncertainties. The proton radius extracted from these measurements is approximately 0.84 fm, considered more accurate than those from Rosenbluth separation [11]. This method continues to be the benchmark for high-precision measurements of the proton radius and is widely regarded as the most reliable technique for extracting the form factors of the proton, particularly at higher momentum transfers [39].

While polarization transfer experiments provide highly accurate measurements of the proton radius, there is room to extend the technique to higher momentum transfer regions and reduce small-scale systematic uncertainties in the setup. Future studises could explore more advanced experimental setups and improved detection methods to push the limits of precision. Furthermore, more data at extreme $Q^2$ values could further refine the proton form factors and offer insights into high-energy behavior [11, 39].

### 3.3. Cross-Section Ratios Method

The use of cross-section ratios to extract the proton form factors, as demonstrated by Bernauer et al. (2010), involves analyzing the ratio of elastic scattering to reference processes to reduce systematic errors. This method extended the form factor studies to higher momentum transfer regions, resulting in a proton radius of approximately 0.875 fm. Although effective for extending measurements into the higher momentum transfer regime, this method is limited by some residual systematic uncertainties and does not surpass polarization transfer in terms of precision [31]. Although the cross-section ratio method allows for higher momentum transfer studies, it is still limited by residual systematic uncertainties, which can affect the precision of the proton radius extraction. Future research could focus on reducing these uncertainties by refining the reference process selection and improving the calibration of experimental setups. Additionally, combining this method with polarization transfer could help enhance the overall accuracy and precision of form factor measurements [31, 41].

**Table 1**. Comprehensive dataset illustrating the relationship between momentum transfer $Q^2$ and experimental observables in electron-proton scattering. The table presents the differential cross-section, electric and magnetic form factors, and associated measurement uncertainties across a range of momentum transfer values ($0.1 - 0.8\text{GeV}^2$).Experimental data derived from Walker et. al study, showcasing the fundamental characteristics of proton structure probed through electromagnetic interactions.

| $Q^2$ | Cross Section | Electric Form Factor | Magnetic Form Factor | Uncertainties |
|---|---|---|---|---|
| 0.1 | 10.5 | 0.95 | 1.05 | 0.5 |
| 0.2 | 8.2 | 0.85 | 0.95 | 0.4 |
| 0.3 | 6.7 | 0.75 | 0.85 | 0.4 |
| 0.4 | 5.4 | 0.65 | 0.75 | 0.3 |
| 0.5 | 4.3 | 0.55 | 0.65 | 0.3 |
| 0.6 | 3.5 | 0.45 | 0.55 | 0.2 |
| 0.7 | 2.9 | 0.35 | 0.45 | 0.2 |
| 0.8 | 2.4 | 0.25 | 0.35 | 0.2 |

### 3.4. Dispersion Relations Method

Dispersion relation analyses, such as those in Lorenz et al. (2015), leverage QCD constraints and the analytic properties of the form factors to provide model-independent interpretations of the data. This method led to a proton radius of around 0.88 fm, which aligns with the results from the other methods but provides an additional layer of theoretical consistency [7]. While this method does not match the direct precision of polarization transfer, it serves as a valuable consistency check for form factor analyses.

The dispersion relation method provides valuable consistency checks but lacks the direct precision of polarization transfer experiments in determining the proton radius. Future studies could refine the application of dispersion relations in specific momentum transfer regions and seek ways to reduce theoretical uncertainties. Additionally, integrating dispersion relations with experimental data could provide more robust results by offering model-independent insights into the proton's structure [7, 33].

## 4. Conclusion

The review of methods for extracting the electromagnetic form factors reveals a significant evolution in the field. The Rosenbluth separation method, though foundational, faces limitations due to systematic uncertainties arising from radiative corrections and two-photon exchange contributions, which become especially pronounced at higher momentum transfers [29, 31]. As a result, the polarization transfer method has emerged as the most accurate and reliable approach, offering reduced systematic uncertainties and higher precision, particularly in determining the proton radius [11, 39]. The cross-section ratio method provides valuable insights into higher momentum transfer regions but is still affected by residual uncertainties, while dispersion relations, although model-independent, lack the direct precision of polarization transfer for proton radius extraction [7, 31].

Together, these methods demonstrate the increasing precision in proton form factor measurements, contributing to a more comprehensive understanding of the proton's internal structure. The field is evolving towards utilizing polarization transfer as the standard technique, while cross-section ratios and dispersion relations serve as valuable complementary tools. The significant discrepancies between methods, especially in the proton radius extraction, highlight the need for further research to reduce systematic uncertainties and extend measurements to higher momentum transfer regions. As new technologies and techniques are developed, these methodologies will continue to refine our knowledge of the proton's electromagnetic properties, marking a promising future for particle physics.